\documentclass[aps,pra,twocolumn,superscriptaddress,showpacs]{revtex4-1}
\usepackage{amsmath}
\usepackage{graphicx}
\newcommand{\be}{\nopagebreak[3]\begin{equation}}
\newcommand{\ee}{\end{equation}}
\newcommand{\ba}{\nopagebreak[3]\begin{eqnarray}}
\newcommand{\ea}{\end{eqnarray}}

\newcommand{\bmult}{\nopagebreak[3]\begin{multline}}
\newcommand{\emult}{\end{multline}}

\begin{document}
\title{Dynamical Decoupling in Optical Fibers:\\Preserving Polarization Qubits
from Birefringent Dephasing}
\author{Bhaskar Roy Bardhan}
\email{broyba1@lsu.edu}
\author{Petr M.\ Anisimov}
\author{Manish K.\ Gupta}
\author{Katherine L.\ Brown}
\affiliation{Department of Physics and Astronomy, Louisiana State University, Baton Rouge, LA 70803}
\author{N.\ Cody Jones}
\affiliation{Edward L. Ginzton Laboratory, Stanford University, Stanford, CA 94305-4088}
\author{Hwang Lee}
\author{Jonathan P.\ Dowling}
\affiliation{Department of Physics and Astronomy, Louisiana State University, Baton Rouge, LA 70803}

\date{\today }

\begin{abstract}

One of the major challenges in quantum computation has been to preserve the coherence of a quantum system against dephasing effects of the environment. The information stored in photon polarization, for example, is  quickly lost due to such dephasing, and it is crucial to preserve the input states when one tries to transmit quantum information encoded in the photons through a communication channel. We propose a dynamical decoupling sequence to protect photonic qubits from dephasing by integrating wave plates into optical fiber at prescribed locations. We simulate random birefringent noise along realistic lengths of optical fiber and study preservation of polarization qubits through such fibers enhanced with Carr-Purcell-Meiboom-Gill (CPMG) dynamical decoupling. This technique can maintain photonic qubit coherence at high fidelity, making a step towards achieving scalable and useful quantum communication with photonic qubits.


\end{abstract}


\maketitle

\section{Introduction}

Qubits are the building blocks for quantum information processing. They are used to store, process and transmit information. Unavoidable couplings between a qubit and its environment introduce uncontrolled evolution of the qubit causing the qubit to lose its ability to exhibit coherent behavior. As a result, the phase of the qubit becomes randomized and the information stored in it is lost. Such decoherence processes stand as a serious obstacle towards achieving scalable quantum information processing.


Dynamical decoupling (DD) is a simple and effective technique that can be used to extend the lifetime of stationary qubit. External pulse sequences are applied to `average out' the qubit-environment interaction by time-reversing the effects of the interaction Hamiltonian. The idea of refocusing the phase diffusion through the application of such repeated external pulses has been extended to a wide variety of DD strategies. Prominent examples of DD schemes are the periodic DD (PDD)~\cite{Viola}, Carr-Purcell DD (CP)~\cite{CP}, Carr-Purcell-Meiboom-Gill (CPMG)~\cite{CPMG}, concatenated DD (CDD)~\cite{Lidar} and Uhrig DD (UDD)~\cite{Uhrig}. In quantum computing, DD has been typically used as an open-loop control scheme to reduce the errors which occur during the evolution of the quantum state.

Photons are a prominent candidate for being mediators in quantum communication processes since they move fast and interact weakly with the environment. Quantum information is often encoded in the polarization or the phase of the photon, or some suitable combination of both degrees of freedom.
Recent developments in the above schemes of DD have motivated us to study whether we can find a suitable DD strategy to preserve polarization photonic qubits against decoherence effects. Previous works have looked at suppression of these effects by using methods including bang-bang decoupling. Wu and Lidar~\cite{Wu} showed that dynamical decoupling could be used for reducing quantum noise in optical fibers. Massar and Popescu presented a method to reduce polarization mode dispersion in optical Þber using controlled polarization rotations~\cite{Popescu}. Entanglement between a single photon and a single trapped atom was reported in Ref.~\cite{Rosenfeld}, which provided a step towards long-distance quantum networking with individual neutral atoms. Damodarakurup \emph{et al.}~experimentally reported on the suppression of polarization decoherence in the ring cavity using bang-bang control of polarization qubits~\cite{Damodarakurup}.

We propose here the application of the CPMG sequence of dynamical decoupling for minimizing the random dephasing in birefringent optical fibers. This sequence has been shown to be robust against a variety of dephasing and control pulse errors~\cite{Morton,Cywinski,Souza,Ajoy}. We apply this sequence to flying polarization qubits in order to extend the useful range of a quantum communication channel. We simulate the CPMG pulses with spatially separated half-wave plates to suppress the dephasing of the input polarization qubits. This will be useful for the BB84 protocol~\cite{BB84} of quantum key distribution, along with applications in the fields like optical quantum computing~\cite{Kok} and quantum networks~\cite{Chaneliere}, with the advantage of having low control overheads. Our proposed scheme helps to improve the range of communication channels without requiring any ancilla qubits and measurement; hence it provides immediate commercial applications for classical telecommunications with light.

This paper is organized as follows. In Sec. II, we discuss the nature of the random fluctuations in a realistic fiber, which are likely to cause dephasing in our polarization qubit. In Sec. III, the key ideas of DD and the motivation for using CPMG are discussed. We provide a detailed description of the proposed application of CPMG to combat such dephasing and preserve the input state through the fiber in Sec. IV. Section V contains our numerical results with explanations. Finally, we conclude with a brief summary of the results, and suggestions for experimentally implementing the DD sequence in optical fibers.

\section{Quantum key distribution in birefringent fibers}

In the BB84 protocol, it is crucial to preserve the input polarized signals against decoherence effects during propagation through the noisy communication channel. Polarization-maintaining (PM) fibers can be used to preserve two orthogonal states of single photon polarization qubit, e.g. photon states with vertical and horizontal polarizations~\cite{VanWiggeren,Rashleigh}. However, the sender needs to randomly choose the input polarized light signals from either the horizontal-vertical basis or the diagonal basis. Therefore, it is necessary to preserve all of these states against unpredictable changes in the polarization state due to dephasing in the fiber~\cite{Shor}.

Due to the random fluctuations in uncontrollable factors like mechanical stress (internal or external)~\cite{Ulrich}, temperature, etc. in the optical fiber, the polarization state of the single photons changes very rapidly as it propagates through the fiber. These effects cause the birefringence $\triangle n=|n_{e}-n_{o}|$ (the subscripts $e$ and $o$ stand for the extraordinary and the ordinary rays) to change randomly along the fiber. The magnitude of the local birefringence at any point along a single mode fiber is typically of the orders of $10^{-4}$ to $10^{-7}$~\cite{VanWiggeren,Rashleigh}. For practical fiber lengths of the order of several hundreds of kilometers, the birefringence in such optical fiber can totally destroy the information stored in the polarization qubits. To combat such large-scale dephasing errors in the fiber, we propose the application of dynamical decoupling implemented with waveplates along the communication channel. As the optical losses are small for the wavelength in the telecommunication band, we restrict our calculations for this wavelength region (around $1550$ nm). We also assume that the waveplates are very thin such that the phase accumulated during the propagation of the photons through the waveplates can be considered negligible.

\section{Dynamical Decoupling}

Dynamical decoupling is an effective method to time-reverse the system-bath interaction by repeatedly applying sequences of sufficiently fast and strong pulses~\cite{BB}. After applying a sequence of these pulses, the evolution of the system is modified such that its interaction with the environment is reduced or eliminated. Since it is difficult to control the states of the environment, the control pulses need to act on the system.

The most general Hamiltonian describing the evolution of a system coupled to a bath can be written as
\begin{equation}
H=H_{\rm S}\otimes I_{\rm B}+I_{\rm S}\otimes H_{\rm B}+H_{\rm I},
\end{equation}
where $H_S$ and $H_B$ are the system and bath Hamiltonians, respectively. When the system is a qubit undergoing dephasing, we can write the interaction Hamiltonian $H_{\rm I}$ as
\begin{equation}
H_{\mathrm{I}} = \sigma_{z} \otimes B_{Z},
\end{equation}
where $\sigma_{z}$ is the Pauli $Z$ spin operator and $B_{Z}$ is a bath operator which couples to the photonic qubit, causing dephasing.  In the limit of weak system-bath coupling, we can invoke the Born-Markov approximation and write the semiclassical interaction Hamiltonian as $H_{\mathrm{I}} = B(t)\sigma_{z}$, where $B(t)$ is scalar function of time.  If we control the qubit with rotations around the $X$-axis on the Bloch sphere, then the complete system Hamiltonian with dephasing is
\begin{equation}
\label{BornMarkov}
H(t) = B(t)\sigma_{z} + f(t)\sigma_{x},
\end{equation}
where $f(t)$ is the time-dependent control field for the dynamical decoupling pulses~\cite{Biercuk}.

The underlying principle of dynamical decoupling is to select a pulse sequence $f(t)$ which causes the integrated time evolution of the interaction Hamiltonian to coherently average to zero~\cite{Haeberlen}. We suppose that
\begin{equation}
f(t) = \frac{\pi}{2} \sum_{k=1}^M\delta(t - t_k),
\end{equation}
or in other words, $f(t)$ consists of instantaneous $\pi$ pulses at prescribed time instants $\{t_1,\ldots,t_M\}$.  We can then move to the interaction picture of the control field $f(t)\sigma_{x}$ and rewrite the system Hamiltonian from Eqn.~(\ref{BornMarkov}) as
\begin{equation}
\label{IntPicture}
\tilde{H}(t) = y(t)B(t)\sigma_{z},
\end{equation}
where $y(t)$ is a switching function which takes on values $\pm 1$, switching polarity at each time $t_k$ corresponding to a $\pi$ pulse in the control sequence. To see the effect of a dynamical decoupling sequence, we calculate the system propagator in the interaction picture as
\begin{align}
\tilde{U}(0,T) = \exp\left(-i\left[\int_0^T y(t)B(t)dt\right] \sigma_{z}\right),
\end{align}
where for simplicity we set $\hbar = 1$. We can drop time-ordering in the propagator because the simplified Hamiltonian in Eqn.~(\ref{IntPicture}) commutes with itself at all times $t \in [0,T]$.

The ideal decoupling sequence would eliminate any dephasing from the system propagator, but since $B(t)$ is in general unknown, decoupling sequences must attempt to make $\int_0^T y(t)B(t)dt \approx 0$, by appropriate choice of pulse locations $\{t_1,\ldots,t_M\}$.  In the simplest example, consider a time-independent $B(t) = B_0$ and single $\pi$ pulse at $t = T/2$.  This is a simplified form of the Hahn spin-echo sequence~\cite{Hahn}, and the system is fully decoupled at time $T$.

A group theoretic understanding of the unitary symmetrization procedure, to eliminate errors up to the first order in the Magnus expansion, is provided in Ref.~\cite{Zanardi}. Typically, the coupling terms contributing to the decoherence errors undergo an effective renormalization. This renormalization transformation can be considered as the cancelation of the terms in the Magnus expansion of the effective Hamiltonian~\cite{Blanes}.  This analysis assumes ideal pulses having zero width, in which case it has been shown that DD sequences can be designed to make higher-order system-bath coupling terms vanish ~\cite{Liu,Biercuk,Lidar}.  However, such ideal, instantaneous pulses are not achievable in actual physical systems, and some researchers have considered finite-width pulses, which decouple to higher order than simple rectanglular pulses~\cite{Uhrig2010,Lidar2011}.

Our aim is to preserve input polarization qubits against dephasing effects in optical fiber, so we intend to implement the most suitable DD sequence to suppress dephasing for a given length of the fiber. The simplest case is the spin-echo sequence~\cite{Hahn}, where it is possible to reduce the pure dephasing due to low-frequency ($\omega<1/\tau$) noise by applying a $\pi$ pulse at time $\tau$/2 during the free evolution. However, this is not very effective in presence of high-frequency noise. Moreover, errors are introduced by imperfections in realistic pulses. These issues can be addressed by repeating the $\pi$ pulses used in the spin-echo technique $N$ times. This choice of pulse sequence is known as the CPMG dynamical decoupling. The CPMG sequence acts as high-pass filter, which effectively filters out the components of the $H_{\rm I}$ varying slowly compared to $\tau$.

Uhrig analytically proposed a non-equidistant sequence of pulses (the UDD sequence~\cite{Uhrig}) which outperforms CPMG when the noise has a sharp, high-frequency cut off~\cite{Bollinger,Jenista,Du}. However, when the spectral density (see Appendix for the description of the decoherence function in terms of spectral density) of the bath has a soft cut-off (long-tail), the CPMG sequence has been shown to give better results than the UDD sequence~\cite{Cywinski,Pasini}. Moreover, the CPMG sequence has been shown to outperform other sequences for an intermediate region where the spectral density is Gaussian~\cite{Abragam}.

The spectral density in our model is taken to be Gaussian with zero mean and this can be considered to be in the intermediate region~\cite{Kurt}. Since we wish to implement CPMG in the optical fiber, it is fairly practical to place suitably oriented half-wave plates along the fiber. This produces the desired polarization rotations and refocuses the quantum state to its original polarization. Another motivation for using such sequence is that CPMG sequence is robust against phase randomization and rotation errors (after every even number of cycles)~\cite{Morton}, particularly when noise processes are dominated by low-frequency components.

\section{Suppressing Birefringent Dephasing with CPMG sequence}

\subsection{Dephasing model}

\begin{figure}
\centering
		\includegraphics[width=\columnwidth,height=3.1in]{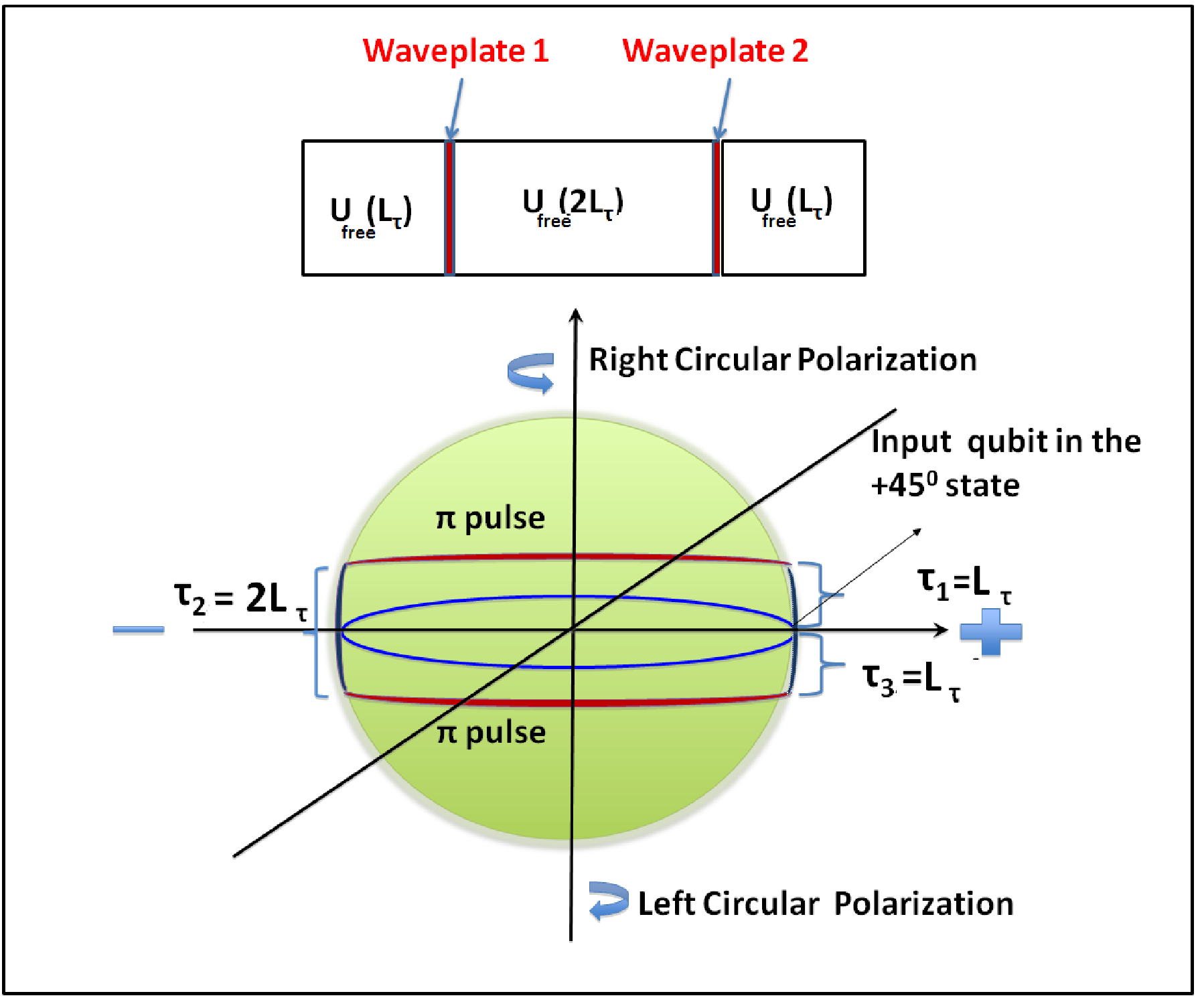}
	\caption{(Color Online) Top : CPMG sequence implemented with half-wave plates in the diagonal basis along the fiber; $U_{\textrm{free}}$'s are the propagators corresponding to the free propagations through the dephasing segments. Bottom : Free propagations and $\pi$ rotations caused by the waveplates for the input qubit in the  $+45^{\circ}$ state are shown on the Bloch sphere.}
	\label{Fig1}
\end{figure}

In a realistic birefringent fiber (typically of lengths 10-1000 km), polarization qubits are likely to experience random effects due to changes in temperature, stress, etc. during propagation. The characteristic length scales for such changes in a single mode fiber may be several meters, i.e. lengths smaller than the fiber beat-lengths~\cite{Galtarossa}. We approximate the communication channel as continuously connected fiber elements, as shown in Fig.~\ref{Fig1}, which have sections of constant birefringence on the order of this length scale. We consider the quantum communication channel provided by a polarization maintaining fiber without dispersion. Assuming that single-photon sources are available, we initialize the qubit in the $+45^{\circ}$ or $-45^{\circ}$ states which can be written as

 \begin{equation}
\left|\psi(0)\right\rangle=\frac{1}{\sqrt{2}}(|H\rangle\pm |V\rangle).
\end{equation}

It should be pointed out here that the following analysis is valid for a general polarization state of single photons, not only the $+45^{\circ}$ or $-45^{\circ}$ states. If we now allow the input photons to propagate freely for a length $L$, then the qubit state becomes
\begin{equation}
|\psi(L)\rangle=\frac{1}{\sqrt{2}}(e^{i\phi_{H}}|H\rangle\pm e^{i\phi_{V}}|V\rangle).
\end{equation}
The phase accumulated by the qubit is given by
 \begin{equation}
 \triangle \phi=\phi_{H}-\phi_{V}=(2\pi/\lambda) \int_{0}^{L}\triangle n(x) dx.
 \end{equation}

 The off-diagonal density matrix element propagates according to

\begin{equation}
\rho_{12}(L)=\rho_{12}(0)\langle\exp{(-i\triangle \phi(x))}\rangle.
\end{equation}


We model the random dephasing by continuously concatenating pieces of fiber with randomly generated lengths $\triangle L$. The total propagation length thus can be split into segments of length $\Delta L$ with constant $\Delta n(x)$. The phase difference for the $i$-th segment is equal to the sum of $(2\pi/\lambda)\Delta L_{i} \Delta n_{i}$. These segments constitute a single phase profile associated with a particular instance of birefringent noise and corresponding changes in the refractive index difference $\triangle n$. Ensemble averaging over profiles gives density matrix for the output state depicting the random dephasing in the fiber.

The correlation function for the refractive index difference at two points $x_{1}$ and $x_{2}$ ($\vert x_{1}-x_{2}\vert$ is less than correlation length) is given by

\begin{equation}
\langle \triangle n(x_{1})\triangle n(x_{1})\rangle=\exp{(-\left[\triangle n^{2}/2\sigma_{\triangle n}^{2}\right])}
\end{equation}

$\triangle n$ being simulated as a Gaussian-distributed zero-mean random process. Estimates of the correlation lengths for a typical optical fiber are given in Ref.~\cite{Galtarossa}.

The Fourier transform $S(k)$ of the correlation function is

\begin{equation}
S(k)=\exp(-\frac{k^{2}\sigma_{n}^{2}}{2}).
\end{equation}

The accumulated phase can be rewritten as
\begin{equation}
\triangle \phi=\left[(2\pi/\lambda)\triangle n\langle \triangle L \rangle\right]\left[\frac{\triangle L}{\langle \triangle L \rangle}\right].
\end{equation}
The two terms in the square brackets are dimensionless. We thus modeled the random birefringent dephasing in a dimensionless manner such that propagation of the input qubit through any given length of the fiber can be simulated just by adjusting the parameters of the randomly generated $\triangle L$.




\subsection{Suppressing birefringent dephasing with CPMG sequence}

The decoherence function (see Appendix) for the dephasing in the fiber (without pulses) can be written as
\begin{equation}
W\left(L\right)=\exp{\left(-\int_{0}^{\infty}\frac{dk}{2\pi} S\left(k\right)\frac{{\sin^{2}(kL)/2}}{k^2}\right)},
\label{eq2}
\end{equation}
where $S(k)$ is the Fourier transform of the autocorrelation function for the random spatial fluctuation of the birefringence~\cite{Cywinski}.

Upon application of a DD pulse sequence with filter function $F(kL)$~\cite{Cywinski,Biercuk} in the space domain, the decoherence function can then be shown to be
\begin{equation}
W\left(L\right)=\exp\left(-\int_{0}^{\infty}\frac{dk}{\pi}S\left(k\right)\frac{F\left(kL\right)}{k^{2}}\right),
\label{eq3}
\end{equation}
where the filter function $F(kL)$ is given by $F(kL)=\frac{1}{2}\left|\sum_{m=0}^{n}\left(-1\right)^{m}\left(e{}^{ikx_{m+1}}-e^{ikx_{m}}\right)\right|^{2}$, corresponding to a pulse sequence having specific set of $x_{m}$ with $x_{0}$=0 and $x_{n+1}=L$.

\begin{figure}
	\centering
	\includegraphics[width=\columnwidth,height=2.8in]{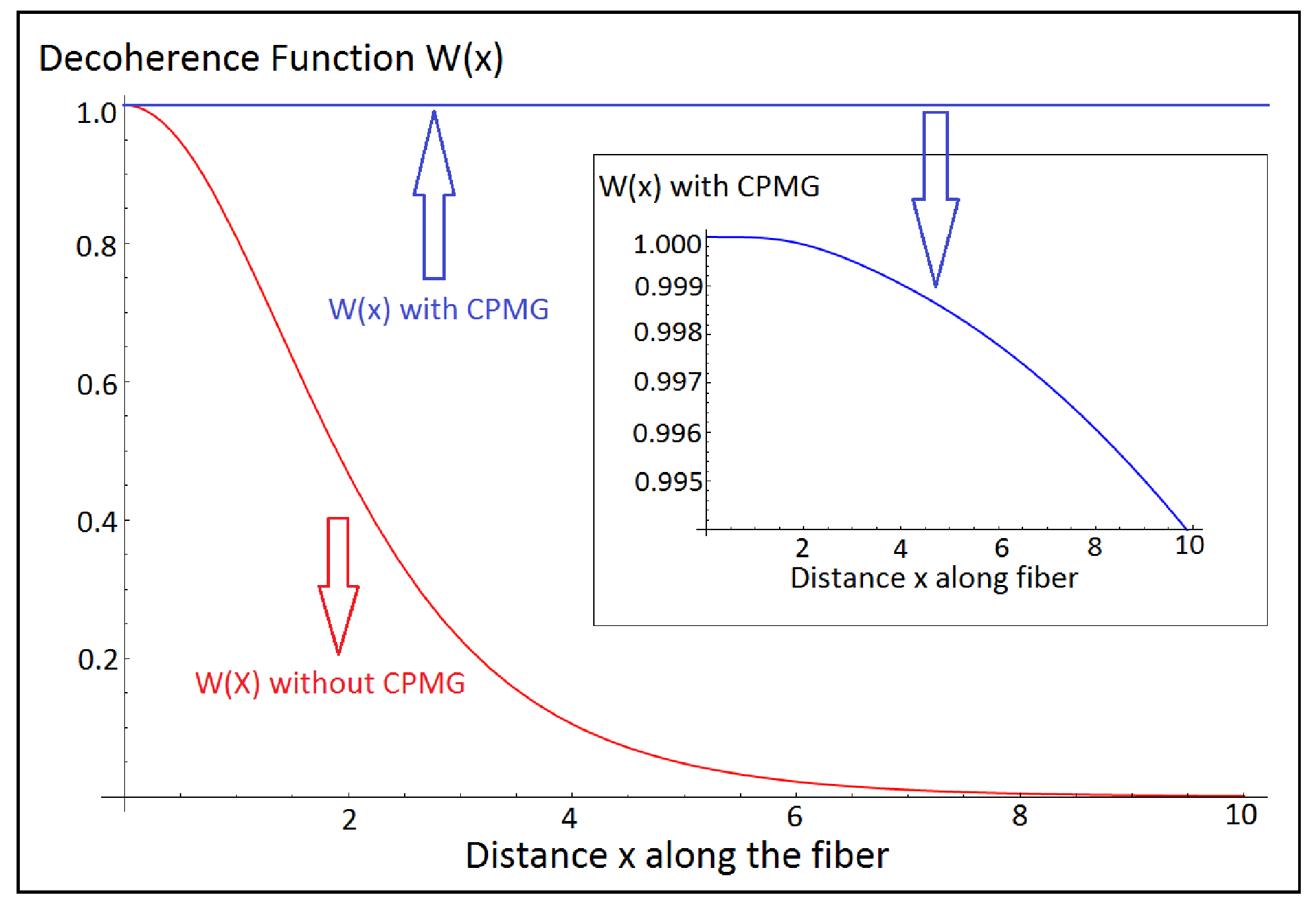}
	\label{Fig2}
	\caption{(Color online) Decoherence function $W (x)$ without CPMG and with CPMG as a function of the distance along the fiber. Inset: $W (x)$ for CPMG with M=2 is shown (zoomed in) with the filter function $F(kL)=8\sin^{4}(kL/8)\sin^{2}(kL/2)/\cos^{2}(kL/4)$ and $S(k)=\exp(-\frac{k^{2}\sigma_{n}^{2}}{2})$. Distance along the fiber is plotted in meters in the figure.}
\end{figure}

Eq.~\eqref{eq3} indicates that the decay rate of the quantum state is determined by the overlap between the spectral density and the filter function $F(kL)$. Moreover, this equation, upon comparison to Eq.~\eqref{eq2}, shows that by choosing suitable $F(kL)$ (and hence DD pulse-sequence), one can expect to reduce decoherence effects introduced during the free propagation of the qubit. We assume that the waveplates are very thin (width of the waveplate $\delta<<L_{\tau}$) such that the phase difference during the propagation of the photons through the waveplates is negligible.


In order to preserve the input polarization qubits in polarization maintaining optical fiber, we simulate the CPMG sequence by placing spatially separated half-wave plates along the fiber. Each waveplate effects a $\pi$ rotation in the qubit state, equivalent to the $\pi$ pulses in typical DD schemes. The multi-pulse CPMG sequence is defined by $x_{k}=L \frac{(k-\frac{1}{2})}{M}$~\cite{CPMG,Cywinski} where the first and the last free propagation periods are half the inter-waveplate separation, effectively refocusing the Bloch vector at the conclusion of the sequence. We take the number of waveplates in one cycle, $M$, to be two. With such sequence the input state remains very well-preserved after propagation through any given length of the fiber.

Each cycle of the two-pulse CPMG is implemented in the following steps (as shown in Fig.~\ref{Fig1}):
\begin{enumerate}
\item Feed a single photon qubit polarized in the $+45^{\circ}$ or $-45^{\circ}$ state into the channel we modeled above.
\item Allow the input qubits to propagate through a segment of the fiber for a length of $L_{\tau}$.
\item Implement the first $\pi$-pulse of the CPMG sequence with half-wave plates in the diagonal basis.
\item Allow a free propagation as before but this time for the length 2$L_{\tau}$.
\item Finally, the second pulse is implemented with a half-wave plate followed by a free propagation of length $L_{\tau}$.
\end{enumerate}

At the end of one cycle of the sequence (having length $4L_{\tau})$, the state of the qubit-fiber system is
$|\psi\left(4L_{\tau}\right)\rangle=\hat{U}|\psi\left(0\right)\rangle$ where the propagator $\hat{U}_{\mathrm{cycle}}$ is given by

\begin{equation}
\hat{U}_{\mathrm{cycle}} =\hat{U}_{\mathrm{free}}(L_{\tau}) \hat{\sigma}_{x}\hat{U}_{\mathrm{free}}(2L_{\tau})\hat{\sigma}_{x}\hat{U}_{\mathrm{free}}(L_{\tau}).
\end{equation}

Here $\hat{\sigma}_{x}$ is the Pauli $X$ operator and $\hat{U}_{\mathrm{free}}(L_{\tau})$ and $\hat{U}_{\mathrm{free}}(L_{\tau})$ are the propagators corresponding to the free propagations along distances $L_\tau$ and $2L_\tau$, respectively.

The CPMG sequence with $2N$ pulses is obtained by repeating the above cycle $N$ times and the propagator for such sequence is $\hat{U}_{\textrm{CPMG}}=\hat{U}_{\mathrm{cycle}}^{N}$ where $\hat{U}_{\mathrm{cycle}}$ is defined above.




%

For the CPMG sequence with $M=2$ pulses in one cycle, the filter function is given by~\cite{Cywinski},
\begin{equation}
F(kL)=8\sin^{4}(kL/8)\sin^{2}(kL/2)/\cos^{2}(kL/4).
\end{equation}
The decoherence functions $W(x)$, without CPMG and with CPMG, are plotted in Fig. 2. We can theoretically predict that CPMG should preserve the coherence of the input polarized qubit for a longer length in the fiber than without the waveplates.


\section{Simulation and Numerical Results}

To characterize the effectiveness of our scheme, we use the fidelity $\mathcal{F}$ between the input state $|\psi_{\rm in}\rangle$  and $\rho_{\rm out}$ as

\begin{equation}
\mathcal{F}=\langle\psi_{\rm in}|\rho_{\rm out}|\psi_{\rm in}\rangle,
\end{equation}
where $\rho_{\rm out}=\frac{1}{n}$ $\displaystyle\sum\limits_{i=1}^n |\psi_{i}\rangle\langle\psi_{i}|$. Here $n$ is the total number of randomly generated phase profiles, corresponding to the propagation operator $\hat{u}_{i}$ so that $|\psi_{i}\rangle=\hat{u}_{i}|\psi_{\rm in}\rangle$ represents the simulated birefringent noise. Therefore, the fidelity being close to one implies that the input state is well-preserved against the dephasing.

\begin{figure}
	\centering
			\includegraphics[width=\columnwidth]{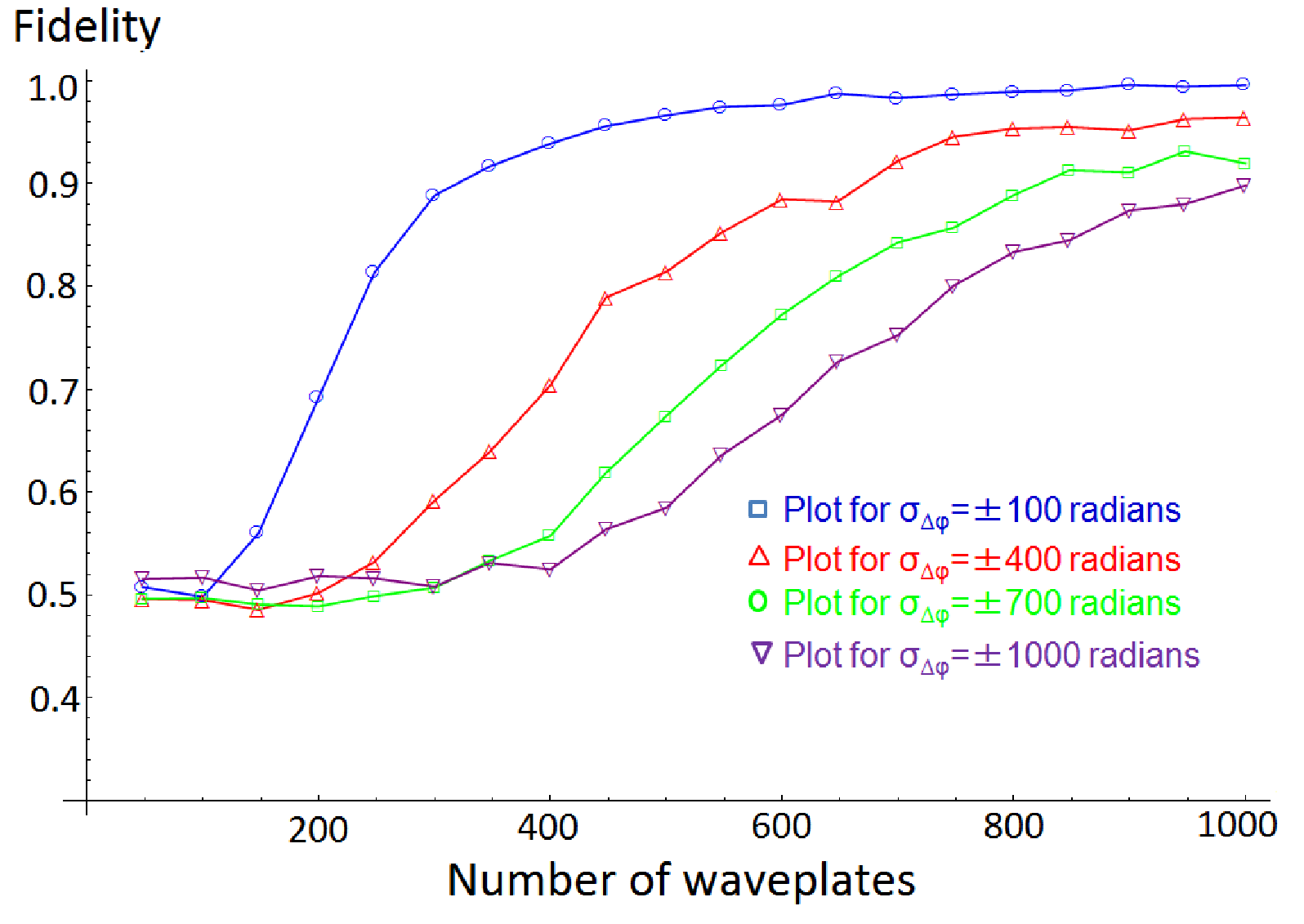}
		\caption{(Color online) Fidelity obtained with CPMG waveplates in the optical fiber is shown with variation of the number of waveplates for different standard deviations of the randomly generated dephasing $\triangle\phi$ and fixed $L=10{\rm km}, \langle \triangle L\rangle=10 {\rm m}$ and $\sigma_{\triangle L}=3 {\rm m}$.}
	\label{Fig3}
\end{figure}

For an arbitrary polarization state $\alpha|H\rangle+\beta|V\rangle$, our calculations show that after $n$ randomly generated phase profiles, the average fidelity between the input and output states is
 \begin{equation}
\mathcal{F}_{Avg}=\Bigg< \cos^{2}(\theta)+\left({|\alpha|}^{2}-{|\beta|}^{2}\right)^{2}\sin^{2}(\theta) \Bigg>.
 \label{Eqa}
 \end{equation}
Here $\theta$ is the total phase introduced by the birefringent fiber as well the waveplates. States which minimize the fidelity are given by $\frac{1}{\sqrt{2}} \left( \left| H \right\rangle + e^{i\phi} \left| V \right\rangle \right)$ and lie on the equator of the Poincar\'{e} sphere. However, Eq.~(\ref{Eqa}) is valid for any general input state  and our simulations, although targeted to improve the fidelity of the states that are useful for the BB84 protocol, hold for any general polarization state, and hence useful for practical implementations of sending a general quantum state through the channel. We observe that the fidelity drastically improves when we used the waveplates even for a large variation of the parameters of the random dephasing $\Delta \phi$. We illustrate this in Fig.~\ref{Fig3}.

\begin{figure}
	\centering
		\includegraphics[width=\columnwidth]{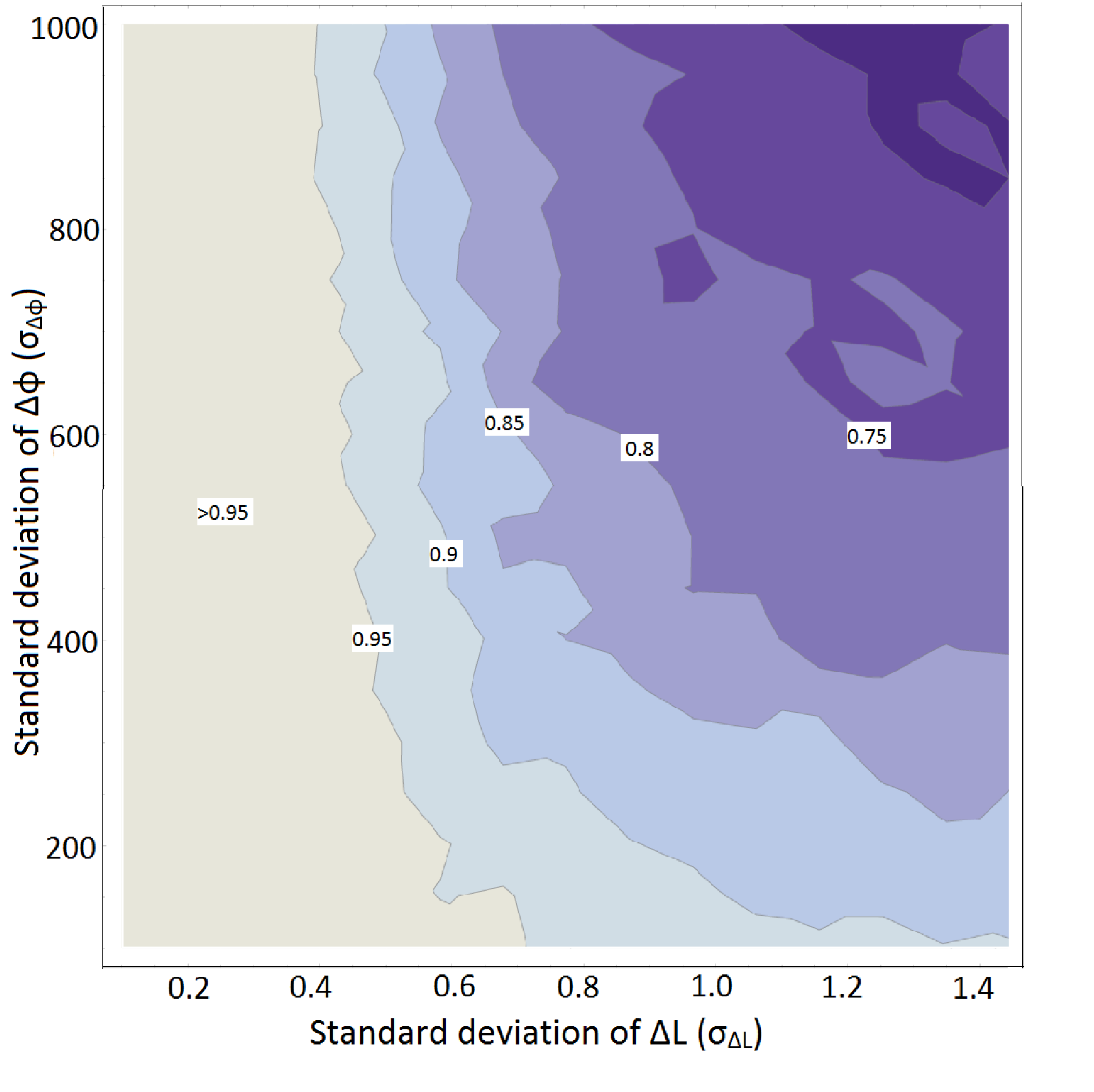}
		\caption{(Color online) Contour plot of the fidelity with the variations of the standard deviations of $\triangle L$ and $\triangle \phi$. Lighter regions show higher values of the fidelity. The simulation is done with fixed $L=10{\rm km}, \langle \triangle L\rangle=10 {\rm m}$.}
	\label{Fig4}
\end{figure}

As described in the Sec. IV, we developed our dephasing model in a dimensionless manner so we can model any realistic fiber length. In Fig.~\ref{Fig4}, the contour plot of the fidelity is shown with respect to the standard deviations $\triangle L$ and $\triangle \phi$. This plot illustrates that high fidelity can be obtained for a reasonably large range of random fluctuations. Therefore, we show how to preserve photonic qubits against dephasing over realistic lengths of optical fiber.

\begin{figure}
	\centering
		\includegraphics[width=\columnwidth]{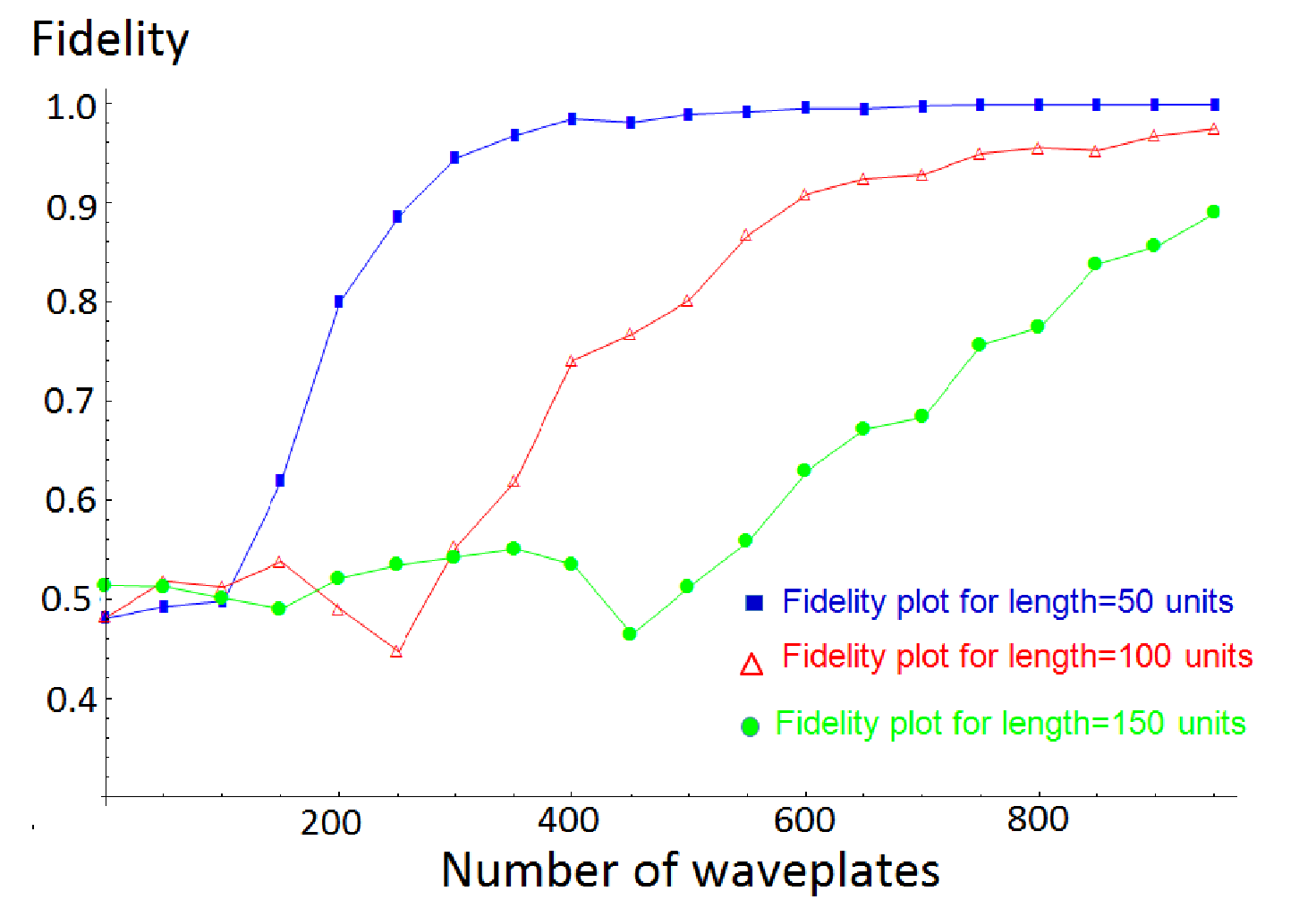}
		\caption{(Color online) Fidelity variation for different fiber lengths with $\langle\triangle L\rangle=10 {\rm m}, \sigma_{\triangle L}=3 {\rm m}$ and $\sigma_{\triangle \phi}=\pm 100$ radians.}
	\label{Fig5}
\end{figure}

The variation of fidelity for different fiber lengths is plotted in Fig.~\ref{Fig5}, which shows that polarization qubits can be preserved up to an excellent fidelity using the CPMG sequence for a wide range of the total fiber length. Moreover, for a given length of the fiber, we can estimate the minimum number of waveplates or the distance between the waveplates required to achieve high fidelity. For instance, for a total fiber length of L=10 km, the estimated number of waveplates from Fig. \ref{Fig3} is 610 to obtain a fidelity of 0.98. Using this, the rough estimate of the inter-waveplate distance  $\L_{\tau}$ is 8.2 m when we considered birefringence fluctuations $\Delta n$ on a length scale of 10 m (which is fairly realistic for the long-distance communication purposes) along the fiber.


\section{Conclusions}

Our dimensionless scheme allows us to tackle the random dephasing using the CPMG sequence for any experimentally viable length of the fiber. Dephasing errors induced by random birefringent noise can be suppressed regardless of magnitude, as long as appropriate waveplate separations are maintained. As we have dealt with noises due to random fluctuations caused by any possible source such as temperature, stress, etc., the prescribed CPMG method can be applied without an experimentalist having a detailed, quantitative knowledge of the decohering environment. Only unitary operations implemented with waveplates are employed; no measurement capabilities nor encoding overheads are required during propagation of the polarization qubits. Moreover, under suitable approximations, this could possibly be extended to address the issue of polarization mode dispersion (PMD) to achieve the desirable high bit-rate for long distance telecommunication in optical fibers ~\cite{Popescu}. To experimentally implement our proposed method to preserve the polarization qubits, several familiar techniques could be suitable depending on the range of fiber lengths one wishes to use. The waveplates may be directly incorporated into the fiber during the manufacturing process. Other methods include writing a Bragg transmission grating periodically into the fiber~\cite{Scalora}, or twisting the fiber in controlled ways causing suitable mechanical stress~\cite{Ulrich}. However, under practical circumstances, we have to consider the finite widths of the waveplates as well rather than the ideal instantaneous pulses and this issue will be addressed in subsequent version of this paper.

We demonstrated in a promising way that it is possible to use CPMG dynamic decoupling to correct for large scale random dephasing errors within the optical fiber. We report for the first time the successful application of the CPMG dynamical decoupling in the birefringent optical fiber which evidently will enhance the range and scope of useful communication through the optical fiber and possibly the coherence time of the qubit. The control overhead in the proposed application of CPMG being reasonably small, we hope that our scheme will reduce the dephasing error while implementing a scalable quantum computing scheme with photonic qubits.

\acknowledgments

We would like to acknowledge support from the AFOSR, IARPA, and NSF. We are thankful to Prof. Kurt Jacobs for helpful discussion.

\section{Appendix}

We model the communication channel as continuously connected fiber elements having sections of constant birefringence. Since we intend to reduce random birefringent noise inside the fiber, the noise spectrum is generated to have a Gaussian probability distribution for such concatenation pieces of the fiber~\cite{Kurt}.

For characterizing the dephasing of the qubits, we define the decoherence function as
\begin{equation}
W(L)\equiv\frac{\left|\langle\rho_{12}(L)\rangle\right|}{\left|\langle\rho_{12}(0)\rangle\right|}=\exp(-\chi(t)),
\end{equation}
When $W(x)=1$, there is no dephasing and $W(x) << 1$ implies that qubit has dephased for the length segment $L$. Here $\chi(t)$ can be expressed through the spectral density $S(k)$ of the noise as~\cite{Cywinski}
\begin{equation}
\label{Appeq}
\chi(t)=\int_{0}^{\infty}\frac{dk}{\pi}S\left(k\right)\frac{F\left(kL\right)}{k^{2}}
\end{equation}
This is the general expression for the function $\chi(t)$ with the filter function $F(kL)$. The filter function encapsulates the influence of the pulse sequence applied.
For the CPMG sequence, we take $F(kL)=8\sin^{4}(kL/4M)\sin^{2}(kL/2)/\cos^{2}(kL/2M)$ \cite{Cywinski} ($M$ is the number of waveplates in one cycle).

Now, to derive an expression showing the effect of the waveplates on the propagation of the qubit state, we consider a CPMG sequence implemented with $M$ waveplates in one cycle. In this case, each of such waveplates causes a $\pi$ rotation about the $x$ axis in the qubit, and this can be written as $\exp(-\textit{i}\frac{\pi}{2}\hat{\sigma}_{x})=-\textit{i}\hat{\sigma}_{x}$. After application of the pulse sequence as illustrated in Fig. 1, the qubit state becomes

\begin{widetext}
\begin{multline}
\mid\psi\left(L\right)\rangle=\mathcal{S}\exp\left[-i\int_{x{}_{N}}^{L}\hat{H}\left(x^{'}\right)dx^{'}\right]\left(-i\hat{\sigma}_{x}\right)\ldots  \\
\exp\left[-i\int_{L{}_{\tau}}^{3L{}_{\tau}}\hat{H}\left(x^{'}\right)dx^{'}\right] \left(-i\hat{\sigma}_{x}\right)
\exp \left[-i\int_{0}^{L{}_{\tau}}\hat{H}\left(x^{'}\right)dx^{'}\right]\mid\psi\left(x=0\right)\rangle,
\end{multline}
\end{widetext}
apart from some constant multiplicative factors in the exponentials. Here $\mathcal{S}$ is the spatial analogue of the time-ordering operator.



\bibliographystyle{h-physrev}

\end{document}